\documentclass[12pt,a4paper]{article}
\usepackage[T1]{fontenc}
\usepackage[latin9]{inputenc}
\usepackage{xspace}
\usepackage{amsmath}
\usepackage{amssymb}
\usepackage{epsfig} 
\usepackage{graphicx}
\usepackage{psfrag,pstricks}
\usepackage[left=2cm,right=2cm,top=2.5cm,bottom=2.5cm]{geometry}

 \def\relic{\Omega_{DM
}}
\def\lsim{\raise0.3ex\hbox{$\;<$\kern-0.75em\raise-1.1ex\hbox{$\sim\;$}}}
\def\gsim{\raise0.3ex\hbox{$\;>$\kern-0.75em\raise-1.1ex\hbox{$\sim\;$}}}

\def\bsg{$b\to s\gamma$}

\def\asusy{a^{\rm SUSY}_\mu}

\DeclareMathAlphabet   {\mathsc}{OT1}{cmr}{m}{sc} 
 
\def\[{\left [} 
\def\]{\right ]} 
\def\({\left (} 
\def\){\right )}

\newcommand{\gappeq}{\mathrel{\rlap {\raise.5ex\hbox{$>$}} 
{\lower.5ex\hbox{$\sim$}}}} 
\newcommand{\lappeq}{\mathrel{\rlap{\raise.5ex\hbox{$<$}} 
{\lower.5ex\hbox{$\sim$}}}}

\newcommand{\bea}{\begin{eqnarray}}
\newcommand{\eea}{\end{eqnarray}}

\newrgbcolor{darkred}{0.7 0 0}
\newrgbcolor{green}{0 1 0}
\newrgbcolor{darkblue}{0 0 0.7}
 \newrgbcolor{purple}{1 0 1}

\begin{document}

\vspace{-1truecm}

\rightline{LPT--Orsay 06/52}
\rightline{ULB-TH/06-30}
\rightline{FTUAM 06/10}
\rightline{IFT-UAM/CSIC-06-33}
\rightline{hep-ph/0607266}
\rightline{July 2006}

\vspace{0.cm}

\begin{center}

{\Large {\bf GLAST versus PAMELA:

A comparison between the detection of gamma rays and positrons 
from neutralino annihilation}}
\vspace{0.3cm}\\

{\large Y. Mambrini$^1$, C. Mu\~noz$^{2,3}$, E. Nezri$^4$
}
\vspace{0.3cm}\\

$^1$ 
Laboratoire de Physique Th\'eorique, 
Universit\'e Paris-Sud, F-91405 Orsay, France
\vspace{0.3cm}\\

$^2$ 
Departamento de F\'isica Te\'orica C-XI,
Universidad Aut\'onoma de Madrid,\\
Cantoblanco, 28049 Madrid, Spain
\vspace{0.3cm}\\

$^3$ 
Instituto de F\'isica Te\'orica C--XVI,
Universidad Aut\'onoma de Madrid,\\
Cantoblanco,
28049 Madrid, Spain
\vspace{0.3cm}\\

$^4$ 
Service de Physique Th\'eorique, CP225, Universit\'e Libre de
Bruxelles,
\\
1050 Bruxelles, Belgium
\vspace{0.3cm}\\

\end{center}

\vspace{-0.5cm}

\abstract{
We study the indirect detection of neutralino dark matter
using positrons and gamma rays from its annihilation in the
galactic halo. Considering the HESS data as the spectrum constituting 
the gamma--ray background, we compare the prospects for the experiments
GLAST and PAMELA in a general supergravity framework with non--universal
scalar and gaugino
masses.
We show that with a boost factor of
about 10, PAMELA will be competitive with GLAST 
for typical NFW cuspy profiles. 
}

\newpage

\vspace{3cm}

\newpage

\tableofcontents

\newpage

\section{Introduction}

The existence of dark matter in the Universe is generally
accepted by the scientific community, though its
nature is still unknown.
One of the most popular candidates, the lightest neutralino,
appears naturally in supersymmetric (SUSY) extensions of the standard model.
It has long been thought that dark matter particles 
could be observed indirectly by detecting the products of
their annihilations. Such products include gamma-rays, 
neutrinos and anti--matter. 
Actually, many experiments have been performed or 
are planned in order to detect indirectly (or directly)
the presence
of such particles in the galactic halo \cite{mireview}.
For example, 
high--energy positron excess in the cosmic rays provides
an opportunity to search for the dark matter signal.
These fluxes will be measured with unprecedented accuracy
by the present space-based experiment PAMELA 
(Payload for Antimatter--Matter Exploration and Light--nuclei
Astrophysics \cite{Stozhkov:2005fw}),
and also by future experiments like 
AMS-02 (Alpha Magnetic Spectrometer \cite{AMS}), designed
to be deployed in the International Space Station
around 2009--10.
The precision measurements of the cosmic positron spectrum to be provided
by PAMELA will be very important to identify
signatures of dark matter in our Galaxy.

On the other hand, the gamma--ray data, richer than the cosmic
particle data, also give us exciting expectations.
The gamma--ray detection experiments
have the advantage of being able to search for point sources
of annihilation radiation. 
In particular, the center of our Galaxy, given its dark matter density,
is one of the most promising sources
of diffuse gamma-rays from dark matter annihilation.
In addition to the controversial EGRET data \cite{EGRET},
atmospheric Cerenkov telescopes like 
HESS \cite{Aharonian:2004wa} and MAGIC \cite{magic}
have observed recently a very bright source in this direction. Actually,
HESS was able to measure in detail the 
gamma--ray spectrum. 
Although fitting the data with a reasonable SUSY model 
seems quite difficult, one can use them as the astrophysical background
for gamma-ray detection \cite{Zaharijas:2006qb}.
Scheduled to begin its five years mission in 2007, 
GLAST (gamma--ray Large Area Space Telescope \cite{GLAST})
will be the most sensitive gamma--ray telescope in
the energy range of interest ($1-300$ GeV).

Both space-based experiments, PAMELA and GLAST, 
will provide us their first results 
around 2009 (before the LHC data).
The aim of this work is to compare the detection prospects
of these two experiments in the framework of supergravity (SUGRA) 
models, taking into account the different astrophysical uncertainties:
boost factor and halo density profile.
In particular, the effect of clumps in the galactic halo may
increase substantially the positron fluxes, and this is parameterised
by
the so-called boost factor. Likewise, cuspy halo profiles may also give
rise to larger gamma--ray fluxes.
Let us finally 
remark that although antiproton fluxes may be in principle as competitive as the positron
ones, we prefer to carry out the analysis of antiprotons
in detail in a future work \cite{Morsellifuture}.

The paper is organized then as follows. In section 2 we will review
the neutralino characteristics as a dark matter candidate in SUGRA
models. In section 3 we will recall the main features of the positron
propagation and detection in the light of the PAMELA experiment.
Section 4 will be devoted to gamma--rays detection and GLAST prospects.
In Section 5 we will compare in detail these two detection modes.
Finally, the conclusions are left for section 6.

\section{Neutralino dark matter}

For the computation of the positron and gamma--ray fluxes 
it is crucial to evaluate the neutralino annihilation cross section.
Of course, for determining this value the theoretical framework
must be established.
In particular, we will work in the context of the
minimal supersymmetric standard model (MSSM). 
Let us recall that in this framework
there are four neutralinos, $\tilde{\chi}^0_i~(i=1,2,3,4)$, 
since they are the physical superpositions of the fermionic partners 
of the neutral
electroweak gauge bosons, called bino ($\tilde{B}^0$) and wino ($\tilde{W}_3^0$), 
and of the fermionic partners of the  
neutral Higgs bosons, called Higgsinos ($\tilde{H}^0_u$, $\tilde{H}_d^0$). 
Thus one can express the lightest neutralino as
\begin{equation}
\tilde{\chi}^0_1 = {Z_{11}} \tilde{B}^0 + {Z_{12}} \tilde{W}_3^0 +
{Z_{13}} \tilde{H}^0_d + {Z_{14}} \tilde{H}^0_u\ .
\label{lneu}
\end{equation}
It is commonly defined that $\tilde{\chi}^0_1$  is mostly gaugino-like 
if $P\equiv \vert {Z_{11}} \vert^2 + \vert {Z_{12}}  \vert^2 > 0.9$, 
Higgsino-like
if $P<0.1$, and mixed otherwise.

\begin{figure}
\centerline{
\epsfig{file=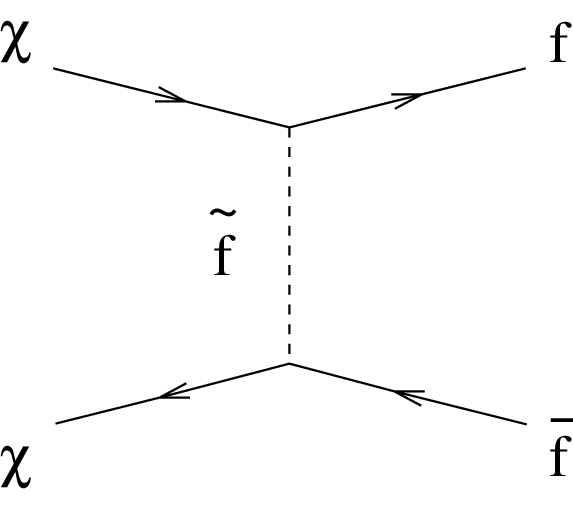,width=0.15\textwidth}\hskip 1cm
    \hspace{1cm}     \epsfig{file=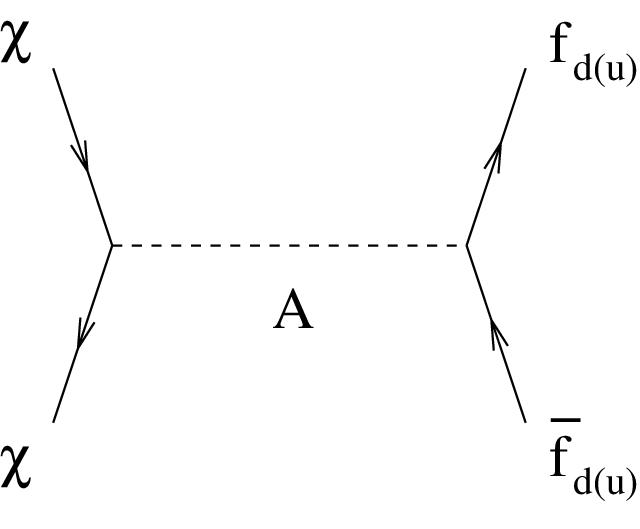,width=0.15\textwidth}\hskip 1cm
       \epsfig{file=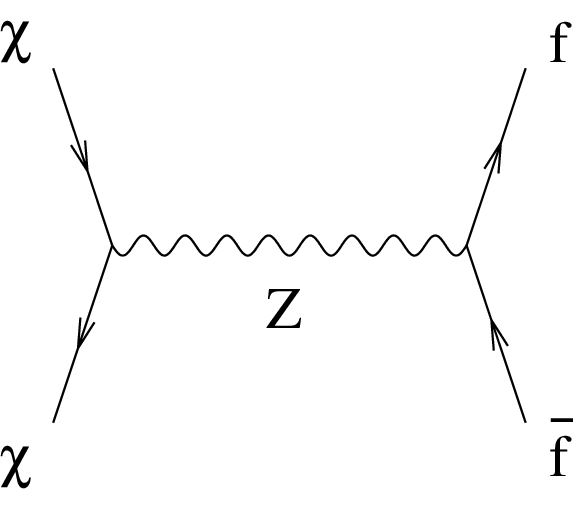,width=0.15\textwidth}\hskip 1cm
      \hspace{1.5cm} 
\epsfig{file=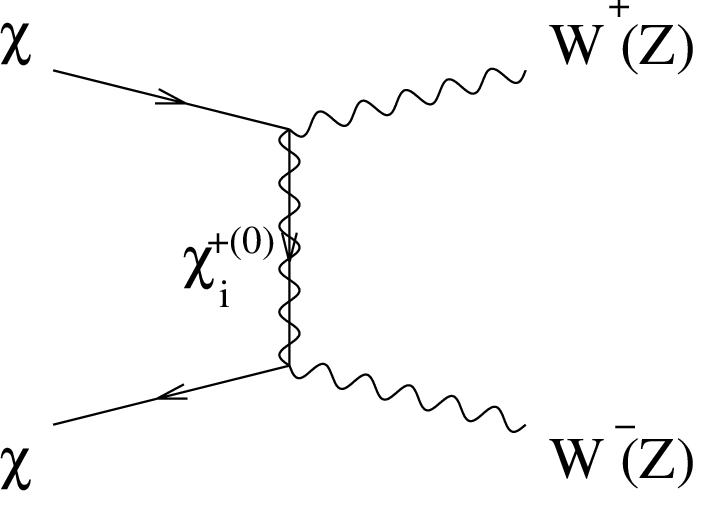,width=0.15\textwidth}\hskip 1cm
        }
 $\propto \frac{m_{\chi} m_f}{m_{\tilde f}^2}Z_{11}^2~~~$
$\propto \frac{m_{\chi}^2}{m_A^2}
\frac{Z_{11}Z_{13,14}}{m_W}  m_{f_d} \tan \beta (\frac{m_{f_u} }{\tan\beta})~ 
~~$
$\propto \frac{m_f m_{\chi}}{m_Z^2} Z_{13,14}^2~$
$\propto \frac{[-Z_{14} V_{21}^* + \sqrt{2} Z_{12} V_{11}^*]^2 (-Z_{13} Z_{31}^* +  
Z_{14} Z_{41}^*)^2}{1+m_{\chi_i^{+(0)}}^2 /
m_{\chi}^2 - m_{W(Z)}^2 / m_{\chi}^2}$
          \caption{{\footnotesize Dominant neutralino annihilation
diagrams. Relevant parts of the amplitudes are shown explicitly
Terms between parenthesis correspond to $f_u$ and $Z$ final states
in second and fourth diagrams. V and Z are chargino and neutralino
mixing
matrices.
}}
        \label{fig:feynmandetail}
\end{figure}

In figure~\ref{fig:feynmandetail} we show the relevant 
Feynman diagrams contributing to neutralino annihilation.
As was 
remarked e.g. 
in Refs.~\cite{Mambrini:2004ke,nonu},
the annihilation cross section
can be significantly enhanced depending on the
SUSY model under consideration.
We will concentrate here on the 
SUGRA scenario, where the soft terms are
determined at the unification 
scale, $M_{GUT}\approx 2\times  10^{16}$ GeV, after SUSY breaking,
and radiative electroweak symmetry breaking is imposed.


\subsection{Supergravity models}

Let us discuss first the minimal supergravity (mSUGRA) scenario,
where the soft terms of
the MSSM are assumed to be universal at  $M_{GUT}$. 
Recall that in mSUGRA
one has only four free parameters:
the soft scalar mass $m$, the soft gaugino mass $M$, 
the soft trilinear coupling $A$, and 
the ratio of the Higgs vacuum
expectation values, 
$\tan\beta= \langle H_u^0\rangle/\langle H_d^0\rangle$.
In addition, the sign of the Higgsino mass parameter, $\mu$,
remains also undetermined by the 
minimization of the Higgs potential.

Since in this scenario the lightest neutralino  $\tilde{\chi}^0_1$ is
mainly bino,
only  ${Z_{11}}$ is large and then the contribution of 
diagrams in Fig.~\ref{fig:feynmandetail} will be generically small,
the first being suppressed by $\tilde{f}$ masses.
As a consequence, for example to reproduce in mSUGRA the 
present experimental accesible regions (EGRET data) 
is not possible \cite{Mambrini:2005vk} for a typical halo model.

However, as discussed in detail in 
Ref.~\cite{Mambrini:2004ke,Mambrini:2005vk} in the context of indirect detection,
the annihilation cross section can be increased 
in different
ways when the structure of mSUGRA for the soft terms is abandoned. 
In particular, it is possible to enhance the annihilation
channels involving the exchange of the CP-odd Higgs, $A$, by reducing the
Higgs mass. 
In addition, it is also possible to  increase the Higgsino components of
the lightest neutralino ${Z_{13,14}}$. 
Thus annihilation channels through Higgs exchange become more important
than in mSUGRA.
This is also the case for $Z$, $\chi_1^\pm$, and  $\tilde{\chi}_1^0$-exchange
channels.
As a consequence, positron and gamma--ray fluxes will be increased.

In particular, the most important effects are produced
by the non-universality of Higgs and gaugino masses.
These can be parameterised, at $M_{GUT}$, as follows
\begin{equation}
  m_{H_{d}}^2=m^{2}(1+\delta_{1})\ , \quad m_{H_{u}}^{2}=m^{2}
  (1+ \delta_{2})\ ,
  \label{Higgsespara}
\end{equation}
and 
\begin{eqnarray}
  M_1=M\ , \quad M_2=M(1+ \delta'_{2})\ ,
  \quad M_3=M(1+ \delta'_{3})
  \ .
  \label{gauginospara}
\end{eqnarray}
We will concentrate in our analysis on the following representative cases:
\begin{eqnarray}
a)\,\, \delta_{1}&=&0\ \,\,\,\,\,\,\,\,,\,\,\,\, \delta_2\ =\ 0
\,\,\,\,\,\,\,\,\,,\,\,\,\, 
\delta'_{2,3}\ =\ 0\ , 
\nonumber\\
b)\,\, \delta_{1}&=&0\ \,\,\,\,\,\,\,\,,\,\,\,\, \delta_2\ =\ 1\
\,\,\,\,\,\,\,\,,\,\,\,\, 
\delta'_{2,3}\ =\ 0\ , 
\nonumber\\
c)\,\, \delta_{1}&=&-1\ \,\,\,\,,\,\,\,\, \delta_2\ =\ 0\
\,\,\,\,\,\,\,\,,\,\,\,\, 
\delta'_{2,3}\ =\ 0\ , 
\nonumber\\
d)\,\, \delta_{1}&=&-1\ \,\,\,\,,\,\,\,\, \delta_2\ =\ 1\
\,\,\,\,\,\,\,\,,\,\,\,\, 
\delta'_{2,3}\ =\ 0\ \,\, , 
\nonumber\\
e)\,\, \delta_{1,2}&=&0\ \,\,\,\,\,\,\,\,,\,\,\,\, \delta'_{2}\ =\ 0
\,\,\,\,\,\,\,\,\,,\,\,\,\, 
\delta'_{3}\ =\ -0.5\ ,
\nonumber\\
f)\,\, \delta_{1,2}&=&0\ \,\,\,\,\,\,\,\,,\,\,\,\, \delta'_{2}\ =\ -0.5
\,\,\,\,\,\,\,\,\,,\,\,\,\, 
\delta'_{3}\ =\ 0\ .
\label{3cases}
\end{eqnarray}
\noindent Case {\it a)} corresponds to mSUGRA with universal soft terms,
cases {\it b)}, {\it c)} and  {\it d)} 
correspond to  non-universal Higgs masses, 
and finally cases {\it e)} and {\it f)} to non-universal gaugino masses.
The cases {\it b)},  {\it c)},  {\it d)}, and {\it e)} were discussed in Ref.~\cite
{Mambrini:2004ke},
and are known to produce gamma--ray fluxes larger than 
in mSUGRA,
whereas case
{\it f)} enhances efficiently annihilation processes because of the
pure wino nature of the LSP.

Let us remark that for the evaluation of the gamma--ray fluxes and 
the positron fluxes at production we use
the last DarkSusy 
released version \cite{darksusynew}. Concerning the computation of the fluxes observed in the solar neighborhood in the case of adiabatically
compressed halos we use our own code.
To solve the renormalization group equations (RGEs) for the soft
SUSY-breaking terms between $M_{GUT}$ and the electroweak scale,
we use the Fortran package SUSPECT \cite{Suspect}.
Concerning the positron spectrum observed by PAMELA, 
we use the annihilation cross section given by DarkSusy at the source,
but solve ourselves the propagation equation and calculation of the flux 
measured on the Earth.




\subsection{Experimental and astrophysical constraints}

We have taken into account in the computation 
several
experimental and astrophysical
bounds. These may produce important constraints 
on the parameter space of SUGRA models, restricting therefore the regions
where the gamma--ray and positron fluxes have to be analyzed.

In particular, concerning the astrophysical constraints, the bounds
$0.1\lsim \relic h^2\lsim 0.3$,
on the  relic neutralino density computation has been considered.
Due to its relevance, the WMAP narrow range
$0.094\lsim\relic h^2\lsim 0.129$ has also been analyzed in detail.

Concerning the experimental constraints,
the lower
bounds on the masses of SUSY particles and on the
lightest Higgs have been implemented, as well as the experimental
bounds on the branching ratio of the \bsg\ process and on 
$\asusy$. 
Note that we are using $\mu > 0$.
We will not consider in the calculation the opposite sign of $\mu$
because this would produce a negative contribution for the $g_{\mu}-2$,
and, as will be discussed below, we are mainly interested in positive
contributions. Recall that the sign of the contribution is basically
given by $\mu M_2$, and that $M$, and therefore $M_2$, can always
be made positive after performing an $U(1)_R$ rotation.
For $\asusy$, we have taken into account the
recent experimental result for the muon
anomalous magnetic moment \cite{g-2}, as well as the most recent
theoretical evaluations of the standard model contributions
\cite{newg2}. It is found that when $e^+e^-$ data
are used the experimental excess in $(g_\mu-2)$ would constrain a
possible SUSY contribution to be
$\asusy=(27.1 \pm 10)\times 10^{-10}$. 
At $2\sigma$ level this implies 
$7.1\times 10^{-10}\lsim\asusy\lsim 47.1\times 10^{-10}$.
It is worth noticing here that when tau data are used a smaller
discrepancy with the experimental measurement is found.
In order not to exclude the latter possibility we will discuss
the relevant value $\asusy=7.1\times 10^{-10}$. 

On the other hand,
the measurements of $B\to X_s\gamma$ decays 
at 
CLEO \cite{cleo} 
and BELLE \cite{belle},
lead to bounds on the branching ratio 
$b\to s\gamma$. In particular we impose on our computation
$2.33\times 10^{-4}\leq BR(b\to s\gamma)\leq 4.15\times 10^{-4}$, where the evaluation 
is carried out using the routine provided by
the program micrOMEGAs \cite{micromegas}.
This program is also used for our evaluation of $\asusy$ and 
relic neutralino density. 
%


\section{The positrons in the Galaxy}

\subsection{The background}

The conventional sources of cosmic rays are believed to be supernovae
and supernovae remnants, pulsars, compact objects in close binary systems, and
stellar winds. Observations of synchrotron emission and $\gamma-$rays
reveal the presence of energetic particles in this objects, thus 
testifying to efficient acceleration processes. Propagation in the
interstellar medium changes the initial composition and spectra
of cosmic ray 
species due to the energy losses (ionization, Coulomb scattering,
bremsstrahlung, inverse Compton scattering, and synchrotron emission),
energy gain (diffusive re-acceleration), and other processes (i.e., 
diffusion and convection by the galactic wind) \cite{Strong:2004de}. 
The destruction of primary
nuclei via spallation gives rise to secondary nuclei and isotopes which
are rare in nature (i.e., Li, Be, B), antiprotons, and pions 
($\pi^{\pm},~\pi^0$) that decay producing secondary $e^{\pm}$ and
$\gamma-$rays. The abundance of stable (Li, Be, B, Sc, Ti, V) and
radioactive ($Be^{10},~Al^{26},~Cl^{36},~Mn^{54}$) secondaries in 
cosmic rays
are used to derive the diffusion coefficient and the halo size
\cite{Moskalenko:1997gh}.
The result of recent simulations agrees with measurements of the 
low--energy positron flux in the cosmic rays \cite{Moskalenko:1997gh}.
The fitting functions for high energy positrons, primary electrons, 
and secondary electrons have been constructed \cite{Baltz1},

\begin{equation}
\Phi_{e^-}^{\mathrm{prim}}(E)=
\frac{0.16 E^{-1.1}}{1+11 E^{0.9} + 3.2 E^{2.15}}
~\mathrm{(GeV^{-1}cm^{-2}s^{-1}sr^{-1})}\ ,
\label{phiemprim}
\end{equation} 

\begin{equation}
\Phi_{e^-}^{\mathrm{sec}}(E)=
\frac{0.70 E^{0.7}}{1+110 E^{1.5} + 580 E^{4.2}}
~\mathrm{(GeV^{-1}cm^{-2}s^{-1}sr^{-1})}\ ,
\label{phiemsec}
\end{equation} 

\begin{equation}
\Phi_{e^+}^{\mathrm{sec}}(E)=
\frac{4.5 E^{0.7}}{1+650 E^{2.3} + 1500 E^{4.2}}
~\mathrm{(GeV^{-1}cm^{-2}s^{-1}sr^{-1})}\ .
\label{phiepsec}
\end{equation}

\subsection{The propagation of a positron} 

Concerning the propagation of positrons, we will use the
"diffusion model" in which the random walk is described by the 
diffusion equation,

\begin{equation}
\frac{\partial}{\partial t} f(E,r)=
K(E) \Delta f(E,r) + \frac{\partial}{\partial E}
[b(E) f(E,r)] + Q(E,r)\ ,
\label{propagation}
\end{equation}
where $f(E,r)$ denotes the number density
of particles per unit of volume and energy and $Q(E,r)$ is
the source (positron injection) term generated by neutralino
annihilation. The flux of positrons with high energy ($E \gg m_e$)
around the Sun is given by:
\begin{equation}
\frac{c}{4 \pi} f(E,r_{\mathrm{\odot}})\ .
\end{equation}
\noindent
Above a few GeV, positron energy losses are dominated by 
synchrotron radiation in the galactic fields and by inverse
Compton scattering on stellar light and on CMB photons.
The energy loss rate $b(E)$ depends on the positron
energy $E$ through : 

\begin{equation}
b(E)=\frac{E^2}{E_0 \tau_E}\ ,
\end{equation}

\noindent
where we have set the energy reference $E_0$ to 1 GeV and the typical
energy loss time is $\tau_E=10^{16}$ s.
We have also assume that the space diffusion coefficient
$K(E)$ is written on the form

\begin{equation}
K(E)=K_0 
\left(
\frac{E}{E_0}
\right)^{\alpha}\ ,
\end{equation}

\noindent
where the diffusion coefficient at 1 GeV is 
$K_0= 3 \times 10^{27} \mathrm{cm^2 s^{-1}}$ with a spectral
index of $\alpha=0.6$ \cite{Baltz1}.

To solve
 Eq.~(\ref{propagation}) we make the hypothesis that the
positron fluxes coming from dark matter annihilation are in equilibrium
in the present universe i.e.,
$\frac{\partial}{\partial t} f(E,r)=0$. Thus Eq.~(\ref{propagation}) simplifies into
\begin{equation}
K_0 \epsilon^{\alpha} \Delta f(\epsilon,r) +
\frac{\partial}{\partial \epsilon}
\left[
\frac{\epsilon^2}{\tau_E}
f(\epsilon,r)
\right]
+Q(\epsilon,r)=0\ ,
\label{eqfinale}
\end{equation}
where we have defined $\epsilon=E/E_0$.

Concerning the boundary condition, we impose the "free escape" one. This
means that the positron density drops to zero on the surface of the
diffusion zone. Moreover, we neglect the positrons coming from
outside our Galaxy. 
The diffusion zone (Galaxy) will be parameterised
by a cylinder of half height $L=3$ kpc and radius $R=20$ kpc.
We will see afterwards that this boundary condition does not affect
too much the result around the Sun.

We can see from Eq.~(\ref{propagation}) that the high--energy 
positrons mainly come from a 
region within few kpc from the solar system. 
In fact, this depends mainly on the parameters $K$ and $b$
appearing there. Indeed, positrons far from the Earth 
lose their energies during the propagation, and consequently
they contribute to the low--energy part of the spectrum.
Using Eq.~(\ref{propagation}) we can easily approximate the distance
in which positrons travel without significant
energy loss:

\begin{equation}
r \approx \sqrt{\frac{K(E)E}{b(E)}}=1.8 \times 
(E/1 \mathrm{GeV})^{-0.2} ~~\mathrm{kpc}\ ,
\label{rmax}
\end{equation}
which gives $r \sim 1.8$ kpc for 1 GeV and $r \sim 0.72$ kpc
for a 100 GeV positron.
This implies that the influence
of the different kinds of dark matter density 
profiles on the positron fluxes is negligible (recall that all 
simulations give rise to the same kind of profile around the solar system).
Thus the
astrophysical 
dependence of the indirect detection of dark matter through
positrons, instead of being similar to the case of the indirect
detection through gamma--rays, looks more like the case of the 
direct detection of
dark matter.

\begin{figure}[!]
    \begin{center}
       \epsfig{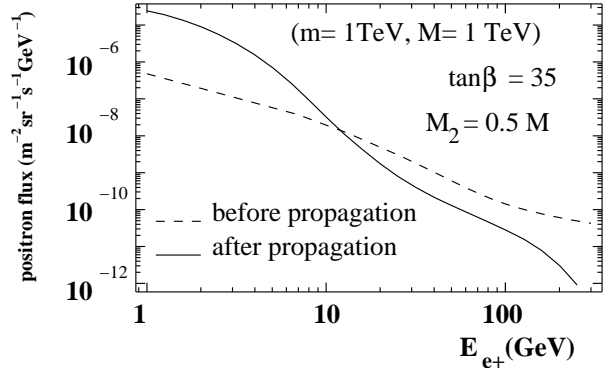}

          \caption{{\footnotesize
The positron flux from neutralino annihilation 
in the non--universal case $f)$
$M_2=0.5M$ with tan$\beta=35$, $\mu=0$, $m=1$ TeV, $M=1$ TeV (corresponding
to a neutralino mass $m_{\chi}=401$) as a function of positron energy 
$E$ before (dashed--line) and after (continuous line) 
propagation in the interstellar
medium.
}}
        \label{fig:Signal}
    \end{center}

\end{figure}

To solve Eq.~(\ref{eqfinale}) we use the treatment given in
Ref.~\cite{Lavalle}, where the solution is expressed with the
Green function $G$ as \footnote{An interesting alternative to solve the propagation
equation is given in \cite{Hisano:2005ec}.}:
\begin{equation}
f(\epsilon,r) = \int_{\epsilon_s=\epsilon}^{\epsilon_s=+\infty} 
d\epsilon_s \int d^3r_s ~ G(\epsilon,\epsilon_s,r,r_s) ~
Q(\epsilon_s,r_s)\ ,
\end{equation}
\noindent
with
\begin{equation}
G(\epsilon,\epsilon_s,r,r_s)=\frac{1-\alpha}
{4 \pi K_0\tau_E(\epsilon^{\alpha-1}-\epsilon_s^{\alpha-1})}
\times
\exp\[\frac{-\rho^2(1-\alpha)}
{4 K_0 \tau_E (\epsilon^{\alpha-1}-\epsilon_s^{\alpha-1})}
\]
\times
V(\epsilon,\epsilon_s,z,z_s)\ ,
\end{equation}
\noindent
and 
\begin{equation}
r=\[x^2 + y^2 + z^2\]^{1/2}~~~~~\ , ~~~~\rho=
\[
(x-x_s)^2 + (y-y_s)^2
\]^{1/2}.
\end{equation}
\noindent
The expression for $V(\epsilon,\epsilon_s,z,z_s)$ is obtained
using the original method of "electrical images" firstly 
introduced by Blatz and Edsjo \cite{Baltz1}
\begin{eqnarray}
V(\epsilon,\epsilon_s,z,z_s)&=&
\sum_{n=1}^{+\infty} \frac{1}{L}
\{
\exp
\[
-\frac{\lambda_n \tau_E(\epsilon^{\alpha-1}-\epsilon_s^{\alpha-1})}
{1-\alpha}
\]\phi_n(z_s)\phi_n(z)\nonumber \\
& &+
\exp
\[
-\frac{\lambda'_n \tau_E(\epsilon^{\alpha-1}-\epsilon_s^{\alpha-1})}
{1-\alpha}
\]\phi'_n(z_s)\phi'_n(z)
\}\ ,
\label{eq:V}
\end{eqnarray}
\noindent 
where $\lambda_n=K_0\times(n-1/2)\frac{\pi}{L}$,
 $\lambda'_n = K_0\times n\frac{\pi}{L}$, and
$\phi_n(z)=\sin{k_n(L-|z|)}$, $\phi'_n(z)=\sin{k'_n(L-z)}$, with
$k_n=1/L(n-1/2)\pi$, $k'_n=1/L n \pi$.
We have checked that (as was already noticed in \cite{Lavalle}) usually
just a few eigenfunctions $\phi_n$, $\phi'_n$, need to be
considered for the sum in Eq.~(\ref{eq:V}) to converge. We illustrate
the effect of the propagation on the spectrum in Fig. \ref{fig:Signal}.
Clearly, this is softened after the propagation through the interstellar
medium.

\subsection{The boost factor}

Recently, many works based on N-body simulations discussed the 
effect of large inhomogeneity (clumps) in the galactic halo. 
In particular, it has been shown 
that dark matter annihilation and detection rates
may be largely enhanced around the solar system \cite{Clumps1,Clumps2,Lavalle}.
The effect is parameterised by a multiplicative factor, the so-called boost factor ($BF$), which
is defined by the ratio of the signal fluxes with 
inhomogeneity and without inhomogeneity,



\begin{equation}
BF=\frac{ \int_{V}d^3x \rho_{clumpy}^2}
{\int_{V}d^3x \rho_{smooth}^2}\simeq \frac{ \int_{V}d^3x
  \rho_{clumpy}^2}{\rho^2_0 V}\ ,
\end{equation}

\noindent
where the region of  integration is taken to be 
$V\sim \mathrm{(a~few~kpc)^3}$ around the solar 
system. 
Obviously, the boost factor is one only when the density  $\rho$
is constant, otherwise it 
is always larger than one. 
Clustering
scenarios suggest a boost factor of 
$\sim 2-5$\footnote{Let us remark however, that in a recent study a model where the
boost factor depends on the energy has been proposed \cite{Lavalle}.
Although a detailed analysis of this possibility is lengthy,
discussions with one of the authors of Ref.~\cite{Lavalle} lead us to
believe
that, due to the characteristics of our positron spectrum, the
results of our analysis would not be crucially affected.}.

\subsection{Profile dependence}

\begin{figure}[!]
    \begin{center}
\epsfig{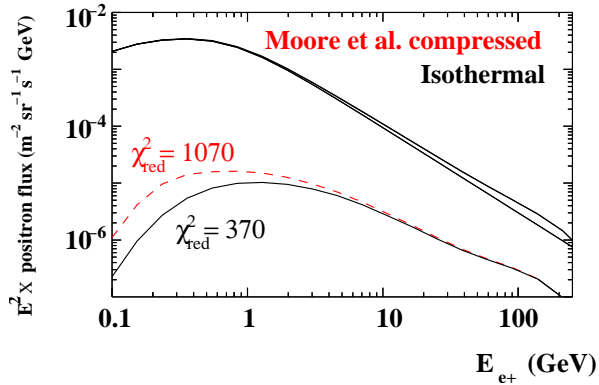}
          \caption{{\footnotesize
Positron flux dependence on the dark matter halo profile. A 
non--cuspy isothermal profile (continuous line) and the very cuspy Moore
et al. profile 
with adiabatic compression (dashed line)
are shown. Black upper curves show the background, and the signal plus background. 
}}
        \label{fig:profiles}
    \end{center}
\label{profiles}
\end{figure}

We show in Fig. \ref{fig:profiles} the influence of the dark matter
distribution on the positron fluxes measured on the Earth. We
have calculated the positron flux in two extreme distributions
around the galactic center:  
a non--divergent isothermal profile and a cuspy one
(Moore et al. profile with adiabatic compression
\cite{Mambrini:2005vk}). Our numerical result clearly agrees with
the approximated result discussed in Eq.~(\ref{rmax}), confirming
that most of the positrons observed would be produced in the
vicinity of the solar system. We remark also that a Moore et al. compressed
profile affects mainly the low--energy positrons in the spectrum.
Indeed, a more cuspy profile will enhance the flux of positrons
coming from
the galactic center, $i.e.$ far away from the solar neighborhood.
They will lose considerable amount of energy during their propagation.
As a consequence, such annihilation process will contribute to 
the low--energy part of the positron spectrum, as we can see in 
Fig. \ref{fig:profiles}.

\subsection{The PAMELA experiment}
\label{PAMELA}

The PAMELA experiment has been launched recently into space.
One of the 
PAMELA primary goals is the measurement of the cosmic
positron spectrum up to 270 GeV \cite{Stozhkov:2005fw}. The geometric
acceptance in the standard trigger configuration is
$\sim$ 20 $\mathrm{cm^2~sr}$. We consider in our analysis
a three years mission and assume Gaussian statistics to
determine the $\chi^2$.

%
%

Let us explain now how we can define that a positron signal
arriving from the galactic halo 
will be distinguished effectively
as a "signal" from the point of view of an experiment. We will
mainly follow in this study the work by Lionetto, Morselli and
Zdravkovic \cite{Lionetto:2005jd} and Hooper and Silk \cite{Hooper:2004bq}.
Let us call $\phi^{susy}$ the signal, 
$\phi^{bkg}$ the background, and $\phi^{tot}=\phi^{susy} + \phi^{bkg}$
the total flux which will be observed by PAMELA. We will divide
our energy range between 1 to 500 GeV in 20 energy bins (N=20)
logarithmically evenly spaced.
For the discrimination between the signal and
the background, we calculate the reduced $\chi^2$, 
$\chi^2_{\mathrm{red}}$:
\begin{equation}
\chi^2_{red}=\frac{1}{N} 
\sum_{n=0}^N
\frac{(E_n^2\phi^{tot}_n-E_n^2\phi^{bkg}_n)^2}{(\sigma_n)^2}\ ,
\label{chi2red}
\end{equation}
where $\sigma_n$ is the error, assuming Gaussian statistic, on
the measured value of the flux multiplied by $E_n^2$.
For a "discovery signal", we request the 
condition\footnote{It is worth noticing here that $\chi^2_{\mathrm{red}}>1$
does not really mean a 1$\sigma$ discovery in our limit.
To be coherent, the important point is to compare two experiments
with the same criterion of discrimination. Thus $\chi^2_{\mathrm{red}}>1$ 
will be the criterion of comparison common to PAMELA and GLAST
in our study.} 
$\chi^2_{red} > 1.$ 

It is easy to check that
this $\sigma_n$, for a given $E_n$, can be expressed as :
%
\begin{equation}
\sigma_n=\sqrt{\frac{E_n^2 (E_n^2\phi^{tot})}{A T}}\ ,
\label{sigma}
\end{equation}
where A is the geometrical factor or acceptance of the experiment
mentioned above,
and $T$ is the exposure time (3 years in our study).
This means that a signal at one $\sigma$ in one energy bin
$E_n$ will give us N hits in the range $N \pm \sqrt{N}$ after T 
seconds (Gaussian law).

\begin{figure}
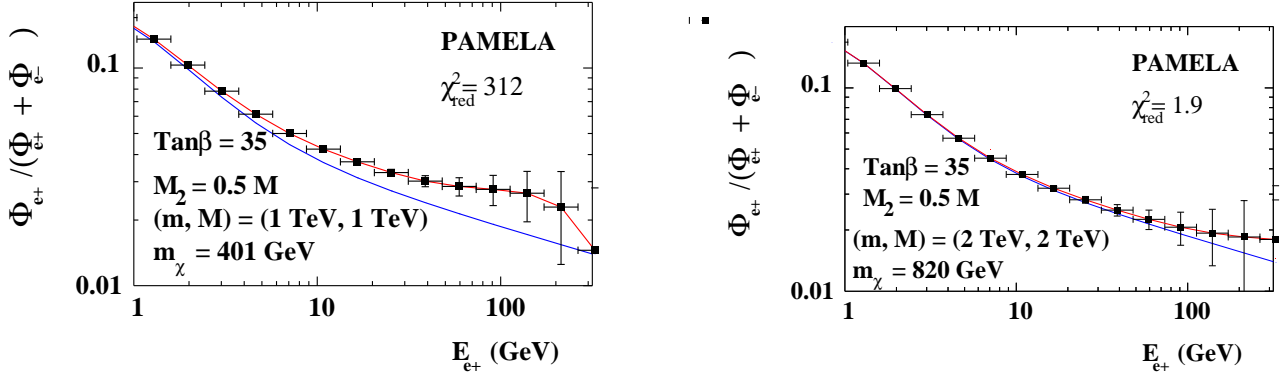

    \begin{center}
       \epsfig{file=NUM2ErrorUs.eps,width=0.45\textwidth}
   \hskip 1cm
       \epsfig{file=NUM2bisErrorUs.eps,width=0.45\textwidth}
          \caption{{\footnotesize
PAMELA expectations for positrons in the non--universal case $f)$
$M_2=0.5M$ with tan$\beta=35$, $A=0$, $m=1$ TeV, $M=1$ TeV (left)
and $m=2$ TeV  $M=2$ TeV (right), corresponding
to a neutralino mass of $m_{\chi}=401$ GeV and $m_{\chi}= 820$ GeV,
respectively. 
The boost factor $BF$ has been set to 5 in an isothermal profile.
The error bars shown are those projected for the PAMELA experiment
after 3 years of observation. The lower (blue) line is the 
background--only prediction.
The reduced $\chi^2$ is 312 in the first case giving 
a signal clearly distinguishable from the background
whereas $\chi^2_{\mathrm red}$ is only 1.9
in the second case and gives a
good fit of the background.
}}
        \label{fig:Signalus}
    \end{center}

\end{figure}

To illustrate the discussion we show in Fig. 
\ref{fig:Signalus} the signal that we can expect for two different points
in the non--universal SUGRA parameter space (in particular for the
case $f)$ $M_2 = 0.5 M$
in Eq.~(\ref{3cases})).
Let us remark that
in addition to the propagation effects, as positrons approach
the solar system, their interactions with the solar wind and
magnetosphere can become important. These effects are called
solar modulation. Alternatively, if one assumes that the effects
of solar modulation are charge-sign independent, their impact 
can be removed considering the ratio of positrons to positrons
plus electrons at a given energy rather than the positron flux
alone. This quantity, the so-called $positron$ $fraction$ is often
used to minimize the uncertainties associated with modelling the impact
of solar modulation.
As we can convert the fluxes to positron fractions by using fluxes of 
background positrons and electrons given by the
Eqs.~(\ref{phiemprim})-(\ref{phiepsec}), we can replace the 
number of events used in Eq.~(\ref{chi2red}) with ratios of
positrons to positrons$+$electrons to reduce the effects of solar 
modulation in our results. Because there are substantially more electrons
than positrons observed, we can assume that there are negligible errors
associated with the electron flux.

\section{The gamma--rays in the Galaxy}
\label{gammaray}

\subsection{The flux}

The spectrum of gamma--rays generated in dark matter annihilation
and coming from a direction forming an angle $\psi$ with respect to 
the galactic center is
\begin{equation}
\Phi_{\gamma}(E_{\gamma}, \psi)
=\sum_i 
\frac{dN_{\gamma}^i}{dE_{\gamma}}
 \langle\sigma_i v\rangle \frac{1}{8 \pi m_{\chi}^2}\int_{line\ of\ sight} \rho^2
\ dl\ ,
\label{Eq:flux}
\end{equation}
where the discrete sum is over all dark matter annihilation
channels,
$dN_{\gamma}^i/dE_{\gamma}$ is the differential gamma--ray yield,
$\langle\sigma_i v\rangle$ is the annihilation cross section averaged 
over its velocity distribution, and $\rho$ is the dark matter density.

\begin{center}
\begin{table}
\centering
\begin{tabular}{|c|ccccc|}
\hline 
&a (kpc)&$\alpha$&$\beta$&$\gamma$
&$\bar{J}(10^{-5} {\rm sr})$  \\
\hline 
NFW & 20 & 1 & 3 & 1 
& $ 1.264 \times  10^4$\\
$\rm{NFW_c}$ & 20 & 0.8 & 2.7 & 1.45 
& $ 1.205  \times 10^7$\\
Moore et al. & 28 & 1.5 & 3 & 1.5 
& $ 5.531  \times 10^6$\\
$\rm{Moore_c}$ & 28 & 0.8 & 2.7 & 1.65 
& $ 5.262  \times 10^8$\\
\hline 
\end{tabular}
\caption{NFW and Moore et al. 
density profiles without 
and with
adiabatic compression ($\rm{NFW}_c$ and $\rm{Moore_c}$ respectively)
with the corresponding parameters, and values of $\bar{J}(\Delta\Omega)$.}
\label{tab}
\end{table}
\end{center}

It is customary to rewrite Eq.~(\ref{Eq:flux}) introducing the
dimensionless quantity $J$ (which depends only on the dark
matter distribution):
\begin{equation}
J(\psi)=\frac{1}{8.5 ~\mathrm{kpc}}
\left(
\frac{1}{0.3 ~\mathrm{GeV/cm^3}}
\right)^2
\int_{line\ of\ sight}\rho^2(r(l,\psi))\ dl\ .
\label{Jbarr}
\end{equation}
After having averaged over a solid angle, $\Delta \Omega$,
the gamma--ray flux can now be expressed as 
\begin{eqnarray}
\Phi_{\gamma}(E_{\gamma})
& = & 
0.94\times 10^{-13}\ \mathrm{cm^{-2}\ s^{-1}\ GeV^{-1}\ sr^{-1}} 
\nonumber \\
&\times & \mbox{}
\sum_i
\frac{dN_{\gamma}^i}{dE_{\gamma}}
\left(
\frac{\langle\sigma_i 
v\rangle}{10^{-29} {\mathrm{cm^3 s^{-1}}}}
\right)
\left(
\frac{100 ~\mathrm{GeV}}{m_{\chi}}
\right)^2
{\overline{J}}(\Delta \Omega) \Delta \Omega \ .
\label{Eq:totflux}
\end{eqnarray}
The value of $\overline{J}(\Delta \Omega) \Delta \Omega$ depends
crucially on the dark matter distribution. 
The different profiles that have been proposed in the literature
can be parameterised as
\begin{equation}
\rho(r)= \frac{\rho_0  [1+(R_0/a)^{\alpha}]^{(\beta-\gamma)/\alpha}  }{(r/R_0)^{\gamma} 
[1+(r/a)^{\alpha}]^{(\beta-\gamma)/\alpha}}\ ,
\label{profile} 
\end{equation}
where $\rho_0$ is the local (solar neighborhood) 
halo density 
and 
$a$ is a characteristic length.
Although we will use $\rho_0=0.3$ GeV/cm$^3$ throughout the paper,
since this is just a scaling factor in the analysis,
modifications to its value can be straightforwardly taken into account
in the results.
N--body simulations suggest a cuspy inner region of dark 
matter halo with a distribution 
where $\gamma$ generally lies in the range 1 (NFW profile \cite{Navarro:1995iw})
to 1.5 (Moore et al. profile \cite{Moore:1999nt})\footnote{For a different 
viewpoint favoring cored distributions, see Ref.~\cite{Salucci}.},
producing a profile with a behaviour $\rho(r) \propto r^{-\gamma}$ at small distances.
Over a solid angle of 
$10^{-5}$ sr, such profiles can lead to 
$\overline{J}(\Delta \Omega)\sim 10^4$
to $10^7$. Moreover, if we take into account the baryon distribution
in the Galaxy, we can predict even more cuspy profiles with
$\gamma$ in the range 1.45 to 1.65
($\overline{J}(\Delta \Omega) \sim 10^7-10^8$) through 
the adiabatic compression process
(see the study of Refs.~\cite{Prada:2004pi,Mambrini:2005vk}).
We summarize the parameters used in our study and the values of
$\bar J$ for each profile in table \ref{tab}.
It is worth noticing here that we are neglecting the effect of
clumpiness, 
even though other studies showed that, depending
upon assumptions on the clumps distribution, in principle an enhacement
of a factor 2 to 10 is possible \cite{Clumps2}. Thus our predictions below for GLAST of the
gamma-ray flux from the galactic center from dark matter pair
annihilations, are conservative in this respect.

\subsection{The background}

HESS \cite{Aharonian:2004wa} has recently measured the 
gamma--ray spectrum from the galactic
center in the range of energy $\sim$ [160 GeV--10 TeV]. The collaboration
 claim that the data are fitted by a power--law 
$F(E) = F_0 ~ E_{\mathrm{TeV}}^{-\alpha}$, with a spectral index 
$\alpha=2.21 \pm 0.09$ and 
$F_0=(2.50 \pm 0.21) \times 10^{-8} ~\mathrm{cm^{-2} ~ s^{-1} ~ TeV^{-1}}$
The data were taken during the second phase of the measurement
(July--August, 2003) with a $\chi^2$ of 0.6 per degree of
freedom. Because of the constant slope
power--law observed by HESS, it results possible but difficult to
conciliate such a spectrum with a signal from dark matter annihilation
\cite{Mambrini:2005vk, Profumo:2005xd}. 
Indeed, final particles (quarks, leptons or gauge bosons) 
produced through annihilations give rise to an spectrum with a continuously 
changing slope. Several astrophysical models have been proposed in order
to match the HESS data \cite{Aharonian:2004jr}.
In the present study we consider the astrophysical background for
gamma--ray detection as the one extrapolated from the HESS data with
a continuous power--law over the energy range of interest
($\approx$ 1 -- 300 GeV).
As was recently underlined in \cite{Zaharijas:2006qb},
GLAST sensitivity is affected by the presence of such an astrophysical 
source.
Note that neutralino masses obtained in our parameter space 
$\lesssim 1$ TeV avoid any conflict with the observations of HESS.

In addition, 
we have also taken into account the EGRET data \cite{EGRET} in our background
at energies below 10 GeV, as they can affect the sensitivity of the analysis.
Indeed, the extrapolation of the gamma--ray fluxes measured
by HESS down to energies as low as 1 GeV is likely to be an underestimation
of the gamma--ray background in the galactic center, as EGRET measurements
are one to two orders of magnitude higher than the HESS extrapolation.
We have decided then to take as our background an interpolation between the HESS
 extrapolation and the EGRET data below 10 GeV to stay as conservative
as possible in evaluating the gamma--ray background.

\subsection{The GLAST experiment}

The space--based gamma--ray telescope GLAST \cite{GLAST} is scheduled 
for launch in 2007 for a five years mission. It 
will perform an all-sky survey covering a large energy range 
($\approx$ 1 -- 300 GeV). With an effective area and angular
resolution on the order of $10^4 ~ \mathrm{cm^2}$ and
$0.1^o$ ($\Delta \Omega \sim 10^{-5}$ sr) respectively,
GLAST will be able to point and analyze the inner center of the Milky
Way ($\sim 7$ pc).

\begin{figure}[!]
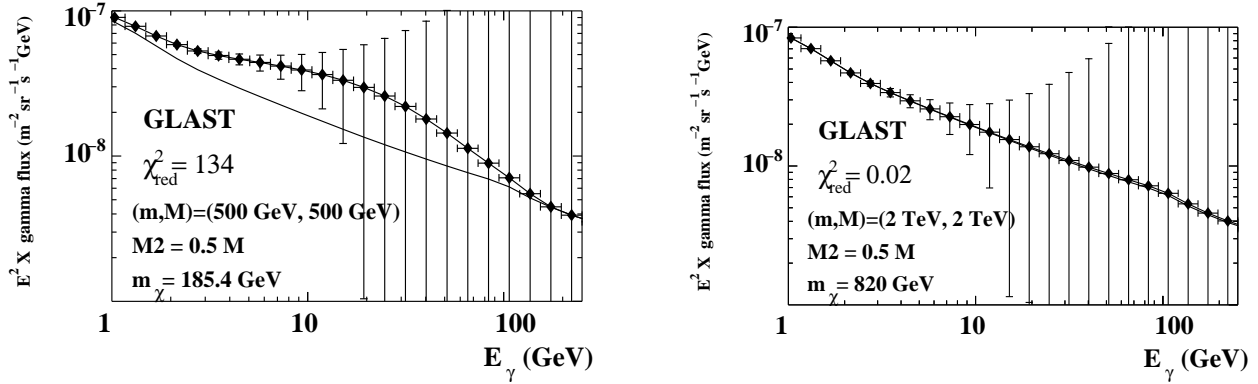

    \begin{center}
       \epsfig{file=NUM2ErrorGLAST.eps,width=0.45\textwidth}
   \hskip 1cm
       \epsfig{file=NUM2bisErrorGLAST.eps,width=0.43\textwidth}

          \caption{{\footnotesize
GLAST expectations for gamma--ray fluxes in the non--universal 
case $f)$ $M_2=0.5M$ with tan$\beta=35$, $A=0$, 
$m=500$ GeV, $M=500$ GeV (left)
and $m=2$ TeV  $M=2$ TeV (right), corresponding
to a neutralino mass of $m_{\chi}=185.4$ GeV and 820 GeV, respectively.
A NFW profile is used in both cases.
The error bars shown are those projected for the GLAST experiment
after 3 years of observation. The lower (blue) line is the 
background--only prediction.
The reduced $\chi^2$ is 134 in the first case giving 
a signal clearly distinguishable from the background,
whereas it is only 0.02 in the second case and gives a
very good fit of the background.
}}
        \label{fig:GLAST}
    \end{center}
\end{figure}

In Fig. \ref{fig:GLAST} we show the ability of GLAST to identify
a signal from dark matter annihilation 
in the non--universal case $f)$ $M_2=0.5 M$ with $\tan \beta=35$
for ($m=500$ GeV, $M= 500$ GeV) and ($m=2$ TeV, $M= 2$ TeV) giving
a neutralino mass of 185.4 GeV and 820 GeV respectively. 
Concerning the requested condition on the $\chi^2_{red}$
for a signal discovery,
we have used an analysis similar to the one considered in the
case of positron detection with the PAMELA experiment in
section \ref{PAMELA}, with a three years mission run.
The error bars shown are projected assuming Gaussian statistic, and
we adopt a power--law background extrapolated from the HESS data.
We find that with a $\chi^2_{red}$ of 134,
GLAST would potentially be able to detect dark matter annihilation
radiation from a neutralino of mass 185.4 GeV, whereas 
with a $\chi^2_{red}$ of 0.02, a signal coming from a neutralino of mass
820 GeV will be below its sensitivity.

\section{Discussion}

\begin{figure}[!]
    \begin{center}
\vskip -1.3truecm
       \epsfig{file=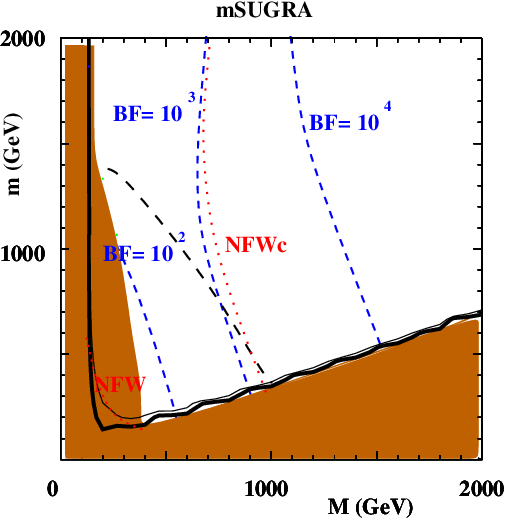,width=0.45\textwidth}
   \hskip 1cm
       \epsfig{file=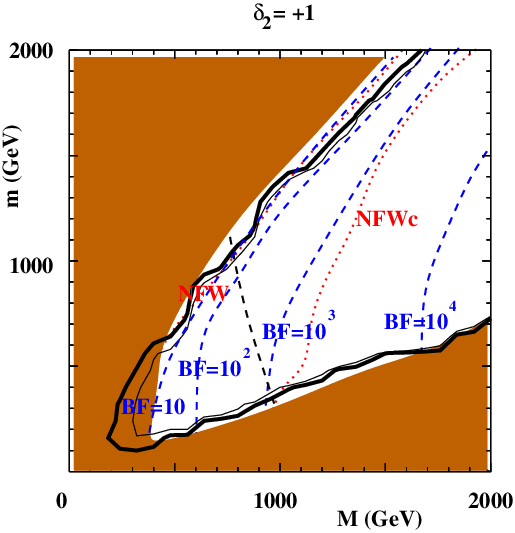,width=0.45\textwidth}
\vskip 0.5cm
       \epsfig{file=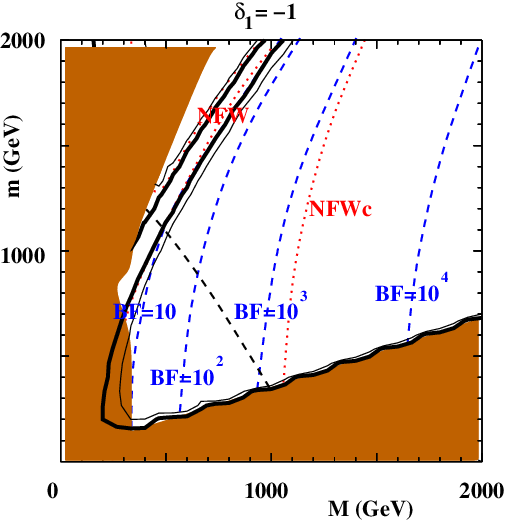,width=0.45\textwidth}
   \hskip 1cm
       \epsfig{file=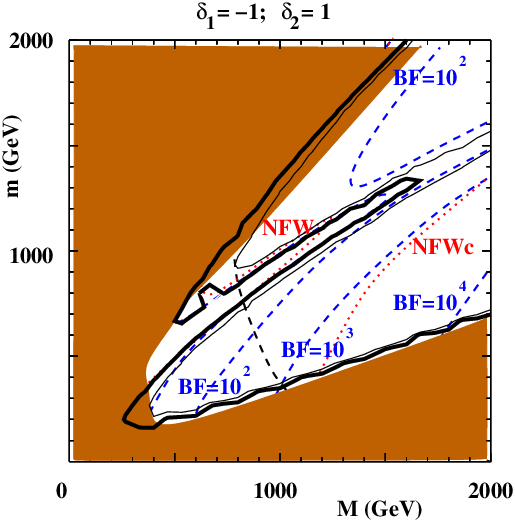,width=0.45\textwidth}
\vskip 0.5cm
       \epsfig{file=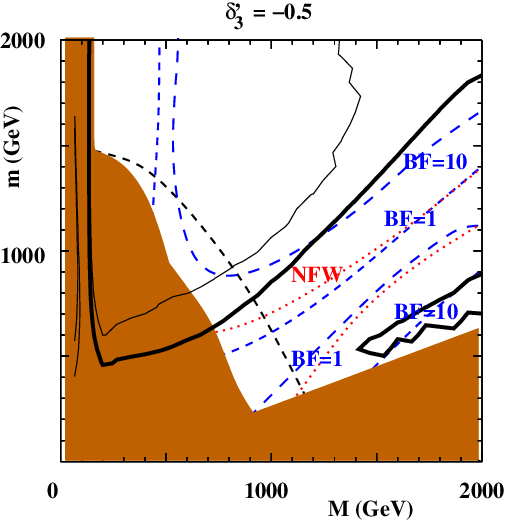,width=0.45\textwidth}
   \hskip 1cm
       \epsfig{file=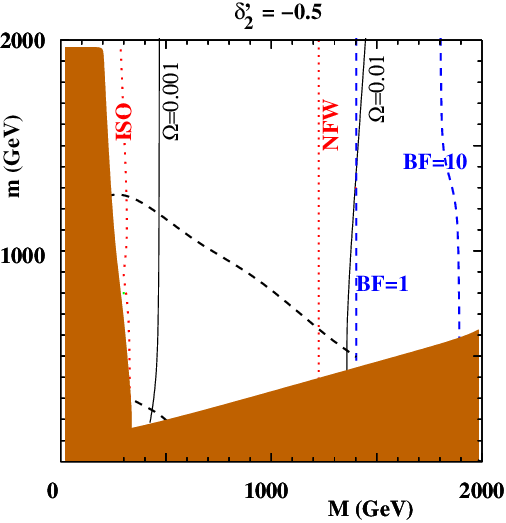,width=0.45\textwidth}

          \caption{{\footnotesize
The isocontour $\chi_{red}^2=1$ for PAMELA (blue--dashed lines)
and GLAST (red--dotted lines) for different kinds of halo profiles
and boost factors, each after three years of observations, 
for tan$\beta=35$, $A_0=0$ and sign($\mu$) $>$ 0.
}}
        \label{fig:Scantb35}
    \end{center}
\end{figure}

We have plotted in Fig.~\ref{fig:Scantb35} the isocontour $\chi_{red}^2=1$
for PAMELA and GLAST, each after three years of observations,
in the six cases 
of Eq.~(\ref{3cases}), and in the SUGRA parameter space (m, M) 
for $\tan\beta = 35$, $A=0$ and $\mu>0$.
We have calculated it 
for boost factors between 1 and $10^4$, and three typical halo profiles,
isothermal (ISO), NFW, and NFW with adiabatic compression (NFW$_c$).
We have taken into account in the computation the astrophysical and
experimental constraints discussed in section~2.2.
In particular, the brown-dark region is excluded by the experimental
constraints and the stau as the lightest SUSY particle.
The region to the right of the black dashed line
corresponds to
$\asusy< 7.1\times10^{-10}$, and would be excluded by
$e^+e^-$ data.
The region between thick and thin solid contours fulfils
$0.1\leq \Omega_{\tilde{\chi}_1^0}h^2\leq 0.3$ (the 
thick contour indicates the WMAP range $0.094<\Omega_{\tilde{\chi}_1^0}h^2<0.129$).
Notice however that we did not rescale the fluxes according to the cosmological
abundance, since 
this procedure would affect photon and positron in the same way and our
work focuses on the 
comparison between the two kinds of detection. As usually discussed in
the literature, 
for points with relic density away from WMAP values, scenarios with
non-thermal production or dilution can be invoked to accomodate the
abundance of neutralinos. 

We clearly see in the figure that the parameter space of mSUGRA will be
reachable only in astrophysical scenarios where the halo profile
is very cuspy, like NFW$_c$ (recall that this profile
is also very similar to a Moore et al. profile without compression), 
or very clumpy with $BF \gtrsim$ 100. 
Notice in this sense 
that for a given $BF$ (halo profile), the region to the right of that
blue-dashed (red-dotted) line 
will correspond to $\chi_{red}^2<1$, and therefore no observation. 
These results are in accordance with a similar
study made in the framework of mSUGRA for the antiproton detection
\cite{Lionetto:2005jd}.
In case {\it b)}, non--universality in 
$m_{H_2}$ ($\delta_2 > 0$) increase the Higgsino fraction of
the neutralino: the annihilation process through $Z$ exchange
is enhanced, permitting observations in NFW--type profiles
or $BF \sim$ 10. Cases {\it c)} and {\it d)} are similar in the sense that
the non-universality for $m_{H_1}$ ($\delta_1 < 0$) open
a broad region of $A-$pole, increasing the gamma ray and
positron fluxes through $\chi \chi \xrightarrow{A} f \overline{f}$,
favored in high tan$\beta$ regimes. Profiles even less cuspy than NFW
and boost factor $\lesssim$ 10 could be sufficient for an observation
by GLAST or PAMELA. The phenomenon is even more spectacular in case 
{\it e)} where the non--universality in the gluino sector, $M_3$ 
($\delta_3' < 0$), acts on $\mu$ and $m_A$ through the RGEs.
As a consequence, the $A-$pole region is broader and allows
observations
even for $BF \sim 1$.
In case {\it f)}, 
the neutralino in mainly Wino and then 
annihilation to positrons is favored (through $W^{\pm}$
final states). In this scenario, a clumpy profile with a 
boost factor of about 1 would be comparable with a cuspy NFW profile.
Unfortunatly, the amount of relic density in either cases is not
sufficient to account for the WMAP results.
As a conclusion, we can say that generically, 
GLAST and PAMELA will have similar discovery prospects
for a NFW profile and $BF \sim 10$, or for a NFW$_c$ profile and $BF \sim 1000$. 
Obviously, a boost factor as large as the latter is unrealistic, and 
in the case of halo models with adiabatic compression PAMELA cannot
really be competitive with GLAST.

\section{Conclusions}

We have studied the indirect detection of neutralino dark matter
using positrons and gamma rays from its annihilation in the
galactic halo. Considering the HESS data as the spectrum constituting 
the gamma--ray background, we have compared the prospects for the 
experiments GLAST and PAMELA in a general SUGRA framework 
with non--universal soft gaugino and scalar masses. 
We have shown that with a clumpiness (boost) 
factor of about 10, PAMELA will be competitive with GLAST 
for a typical NFW cuspy profile. However, 
for the case of a NFW compressed profile, PAMELA cannot be competitive
with GLAST since an unrealistic clumpiness factor of  about
1000 would be necessary. 
For the future, it would be interesting to compile all the detection modes
(antiproton, deuteron, positron and gamma rays) in order 
to carry out a complete
analysis of the parameter space in a general SUGRA framework
\cite{Morsellifuture}.

\vspace*{1cm}
\noindent{\bf Acknowledgements}

We thank Igor Moskalenko and Andrew Strong for their friendly collaboration
during this work.
We are also grateful to Pierre Salati and Julien Lavalle for having shared
with us their knowledge about the propagation of positrons
in our Galaxy. 
The authors want specially to thank warmly Aldo 
Morselli and Andrea Lionetto for their help and discussions on theoretical
and experimental issues concerning GLAST and PAMELA. 


The work of
Y.M. was sponsored by the PAI programm PICASSO under contract
PAI--10825VF. 
The work of C.M. was supported 
in part by the Spanish DGI of the 
MEC under Proyectos Nacionales BFM2003-01266 and FPA2003-04597
and under Acci\'on Integrada Hispano-Francesa HF-2005-0005;
by the Comunidad de Madrid under Ayudas de I+D S-0505/ESP-0346;
and also by the European Union under the RTN program  
MRTN-CT-2004-503369, and under the ENTApP Network of the ILIAS project
RII3-CT-2004-506222.
The work of E.N. was supported by the I.I.S.N. and the 
Belgian Federal Science Policy (return grant and IAP 5/27).

\nocite{}
\bibliography{bmn}

\begin{thebibliography}{99}





\bibitem{mireview} For recent reviews, see:
C.~Munoz,
`Dark matter detection in the light of recent experimental results',
  {\it Int. J. Mod. Phys.}{\bf A19} (2004) 3093
  [arXiv:hep-ph/0309346];  
G.~Bertone, D.~Hooper and J.~Silk,
`Particle dark matter: Evidence, candidates and constraints',
  {\it Phys. Rept.} {\bf 405} (2005) 279
  [arXiv:hep-ph/0404175].


\bibitem{Stozhkov:2005fw}
  Y.~I.~Stozhkov et al. [PAMELA Collaboration],
`About separation of hadron and electromagnetic cascades in the PAMELA
calorimeter',
{\it Int. J. Mod. Phys.} {\bf A20} (2005) 6745;
 P.~Picozza and A.~Morselli, `The science of PAMELA space mission,' 
arXiv:astro-ph/0604207.

\bibitem{AMS}
F.~Barao  [AMS-02 Collaboration], 
`AMS: Alpha Magnetic Spectrometer on the International Space Station',
{\it Nucl. Instrum. Meth.} {\bf A535} (2004) 134.




\bibitem{EGRET} 
S.~D. Hunger et al. [EGRET Collaboration], `EGRET observations of the
diffuse gamma-ray emission from the galactic plane',
{\it Astrophys. J.} {\bf 481} (1997) 205; 
H.~A. Mayer-Hasselwander et al., 
`High-energy gamma-ray emission from the galactic center'
{\it Astron. \& Astrophys.} {\bf 335} (1998) 161.



\bibitem{Aharonian:2004wa}
  F.~Aharonian et al. [HESS Collaboration],
 `Very high energy gamma--rays from the direction of Sagittarius A*',
 {\it Astron. Astrophys.} {\bf 425} (2004) L13
  [arXiv:astro-ph/0408145].


\bibitem{magic}
  J. Albert et al. [MAGIC Collaboration],
 `Observation of gamma rays from the galactic center with the MAGIC telescope',
 {\it Astrophys. J.} {\bf 638} (2006) L101
  [arXiv:astro-ph/0512469].




\bibitem{Zaharijas:2006qb}
G.~Zaharijas and D.~Hooper,
`Challenges in detecting gamma--rays from dark matter annihilations in the
galactic center',
  arXiv:astro-ph/0603540.


\bibitem{GLAST} N. Gehrels and P. Michelson,
`GLAST: the next generation high-energy gamma--ray astronomy mission',
{\it Astropart. Phys.} {\bf 11} (1999) 277.



\bibitem{Morsellifuture} A.~Lionetto, Y.~Mambrini, A.~Morselli, C.~Munoz,
E.~Nezri, in preparation.


\bibitem{Mambrini:2004ke}
Y.~Mambrini and C.~Mu\~noz,
`Gamma--ray detection from neutralino annihilation in non-universal SUGRA
scenarios', 
{\it Astrop. Phys.} {\bf 24} (2005) 208 [arXiv:hep-ph/0407158];
Y.~Mambrini and C.~Mu\~noz,
`A comparison between direct and indirect dark matter search',
{\it JCAP} {\bf 10} (2004) 003
[arXiv:hep-ph/0407352].


\bibitem{nonu}
  H.~Baer, A.~Mustafayev, S.~Profumo, A.~Belyaev and X.~Tata,
`Direct, indirect and collider detection of neutralino dark matter in
SUSY models with non-universal Higgs masses',
{\it JHEP} {\bf 07} (2005) 065
  [arXiv:hep-ph/0504001];
 H.~Baer, A.~Mustafayev, E.~K.~Park, S.~Profumo and X.~Tata,
`Mixed higgsino dark matter from a reduced SU(3) gaugino mass:
Consequences for dark matter and collider searches',
  arXiv:hep-ph/0603197;
H.~Baer, A.~Mustafayev, E.~K.~Park and S.~Profumo,
`Mixed Wino dark matter: Consequences for direct, indirect and 
collider detection',
  {\it JHEP} {\bf 07} (2005) 046
  [arXiv:hep-ph/0505227];
V.~Bertin, E.~Nezri and J.~Orloff,
 `Neutralino dark matter beyond CMSSM universality,'
 {\it J. High Energy Phys.} {\bf 02} (2003) 046
 [arXiv:hep-ph/0210034];
 A.~Birkedal-Hansen and B.~D.~Nelson,
 `Relic neutralino densities and detection rates with nonuniversal 
 gaugino masses,'
 {\it Phys. Rev.} {\bf D67} (2003) 095006
 [arXiv:hep-ph/0211071].


\bibitem{Mambrini:2005vk}
Y.~Mambrini, C.~Munoz, E.~Nezri and F.~Prada,
`Adiabatic compression and indirect detection of supersymmetric dark matter',
{\it JCAP} {\bf 01} (2006) 010 [arXiv:hep-ph/0506204].


\bibitem{darksusynew}
P. Gondolo, J. Edsjo, P. Ullio, L. Bergstrom, M. Schelke and E.~A. Baltz,
`DarkSUSY: Computing supersymmetric
dark matter properties numerically',
arXiv:astro-ph/0406204;
See also the web page
http://www.physto.se/\char126edsjo/darksusy


\bibitem{Suspect}
A. Djouadi, J.~L. Kneur and G. Moultaka,
`SuSpect: a Fortran code for the supersymmetric and Higgs particle
spectrum in the MSSM', arXiv:hep-ph/0211331; See also the web page
 http://www.lpm.univ-montp2.fr:6714/\char126kneur/suspect.html.



\bibitem{g-2} G.~W.~Bennett et al. [Muon g-2 Collaboration],
`Measurement of the negative muon anomalous magnetic moment to
0.7-ppm', {\it Phys. Rev. Lett.} {\bf 92} (2004) 161802
[arXiv:hep-ex/0401008].
  

\bibitem{newg2}
  M.~Davier, S.~Eidelman, A.~Hocker and Z.~Zhang,
  `Updated estimate of the muon magnetic moment using revised results
  from $e^+ e^-$ annihilation',
  {\it Eur. Phys. J.} {\bf C31} (2003) 503
  [arXiv:hep-ph/0308213];
K.~Hagiwara, A.~D.~Martin, D.~Nomura and T.~Teubner,
  `Predictions for $g-2$ of the muon and $\alpha_{QED}(M_Z^2)$',
{\it Phys. Rev.} {\bf D69} (2004) 093003
[arXiv:hep-ph/0312250];
J.~F.~de Troconiz and F.~J.~Yndurain,
  `The hadronic contributions to the anomalous magnetic moment of the
  muon', {\it Phys. Rev.} {\bf D71} (2005) 073008
  [arXiv:hep-ph/0402285].




\bibitem{cleo} 
 S. Chen et al. [CLEO Collaboration], 
`Branching fraction and photon energy spectrum for
$b\to s\gamma$',
{\it Phys. Rev. Lett.} {\bf 87} (2001) 251807
[arXiv:hep-ex/0108032].

\bibitem{belle}
H. Tajima [BELLE Collaboration]
`Belle B physics results', 
{\it Int. J. Mod. Phys.} {\bf A17} (2002) 2967
[arXiv:hep-ex/0111037];
See also the web page http://wwwlapp.in2p3.fr/lapth/micromegas


\bibitem{micromegas} G. Belanger, F. Boudjema, A. Pukhov and
A. Semenov,
`micrOMEGAs: a program for calculating the relic density in the
MSSM',
{\it Comput. Phys. Commun.} {\bf 149} (2002) 103 [arXiv:hep-ph/0112278];
G. Belanger, F. Boudjema, A. Pukhov and A.G. Semenov, 
`MicrOMEGAs: Version 1.3',
[arXiv:hep-ph/0405253]; 
G.~Belanger, F.~Boudjema, A.~Pukhov and A.~Semenov,
``micrOMEGAs2.0: A program to calculate the relic density of dark matter in a
generic model,'' arXiv:hep-ph/0607059.


\bibitem{Strong:2004de}
  A.~W.~Strong, I.~V.~Moskalenko and O.~Reimer,
 `Diffuse Galactic continuum gamma--rays. A model compatible with EGRET data
  and cosmic-ray measurements',
  {\it Astrophys. J.}  {\bf 613} (2004) 962
  [arXiv:astro-ph/0406254].






\bibitem{Moskalenko:1997gh}
  I.~V.~Moskalenko and A.~W.~Strong,
`Production and propagation of cosmic-ray positrons and electrons',
  {\it Astrophys. J.}  {\bf 493} (1998) 694
  [arXiv:astro-ph/9710124].



\bibitem{Baltz1}
E.~A.~Baltz and J.~Edsjo, 
`Positron propagation and fluxes from neutralino annihilation in the halo',
Phys. Rev. {\bf D59} (1999) 023511
[arXiv:astro-ph/9808243].


\bibitem{Lavalle}
J.~Lavalle, J.~Pochon, P.~Salati and R.~Taillet, 
`Clumpiness of Dark Matter and Positron Annihilation Signal: 
Computing the odds of the Galactic Lottery', arXiv:astro-ph/0603796.


\bibitem{Hisano:2005ec} 
J.~Hisano, S.~Matsumoto, O.~Saito and M.~Senami,
``Heavy Wino-like neutralino dark matter annihilation into antiparticles,''
  Phys.\ Rev.\ D {\bf 73} (2006) 055004
  [arXiv:hep-ph/0511118]; 
M.~Asano, S.~Matsumoto, N.~Okada and Y.~Okada,
   ``Cosmic positron signature from dark matter in the littlest Higgs model
with T-parity,''
  arXiv:hep-ph/0602157.



\bibitem{Clumps1}
J.~Silk and A.~Stebbins, 
`Clumpy cold dark matter',
{\it Astrophys. J.} {\bf 411} (1993) 439.

\bibitem{Clumps2}
L.~Bergstrom, J.~Edsjo, P.~Gondolo and P. Ullio, 
`Clumpy neutralino dark matter',
Phys. Rev. D {\bf 59}
(1999) 043506 [astro-ph/9806072].







\bibitem{Lionetto:2005jd}
  A.~M.~Lionetto, A.~Morselli and V.~Zdravkovic,
  `Uncertainties of cosmic ray spectra and detectability of 
   antiproton mSUGRA contributions with PAMELA',
  JCAP {\bf 09} (2005) 010
  [arXiv:astro-ph/0502406].


\bibitem{Hooper:2004bq}
  D.~Hooper and J.~Silk,
  `Searching for dark matter with future cosmic positron experiments',
  {\it Phys. Rev.} {\bf D71} (2005) 083503
  [arXiv:hep-ph/0409104].




\bibitem{Navarro:1995iw}
  J.~F.~Navarro, C.~S.~Frenk and S.~D.~M.~White,
`The Structure of Cold Dark Matter Halos',
{\it Astrophys. J.}  {\bf 462} (1996) 563
  [arXiv:astro-ph/9508025];
 J.~F.~Navarro, C.~S.~Frenk and S.~D.~M.~White,
`A Universal Density Profile from Hierarchical Clustering',  
{\it Astrophys. J.}  {\bf 490} (1997) 493
  [arXiv:astro-ph/9611107].

\bibitem{Moore:1999nt}
  B.~Moore, S.~Ghigna, F.~Governato, G.~Lake, T.~Quinn, J.~Stadel and P.~Tozzi,
`Dark matter substructure within galactic halos',
{\it Astrophys. J.}  {\bf 524} (1999) L19.


\bibitem{Salucci}
G. Gentile, P. Salucci, U. Klein, D. Vergani and P. Kalberla,
MNRAS 351 (2004) 903 [arXiv:astro-ph/0403154].


\bibitem{Prada:2004pi}
  F.~Prada, A.~Klypin, J.~Flix, M.~Martinez and E.~Simonneau,
`Astrophysical inputs on the SUSY dark matter annihilation  detectability',
  arXiv:astro-ph/0401512;
  G.~Bertone and D.~Merritt,
`Dark matter dynamics and indirect detection',
  {\it Mod. Phys. Lett.} {\bf A20} (2005) 1021
  [arXiv:astro-ph/0504422];
 G.~Bertone and D.~Merritt,
`Time-Dependent Models for Dark Matter at the Galactic Center',
  {\it Phys. Rev.} {\bf D72} (2005) 103502
  [arXiv:astro-ph/0501555].
  E.~Athanassoula, F.~S.~Ling and E.~Nezri,
`Halo geometry and dark matter annihilation signal', 
  {\it Phys. Rev.} {\bf D72}, 083503 (2005)
  [arXiv:astro-ph/0504631].



\bibitem{Profumo:2005xd}
  S.~Profumo,
`TeV gamma--rays and the largest masses and annihilation cross sections
of
neutralino dark matter',
  {\it Phys. Rev.} {\bf D72} (2005) 103521
  [arXiv:astro-ph/0508628];
 D.~Hooper, I.~de la Calle Perez, J.~Silk, F.~Ferrer and S.~Sarkar,
`Have atmospheric Cerenkov telescopes observed dark matter?',
  {\it JCAP} {\bf 09} (2004) 002
  [arXiv:astro-ph/0404205].


\bibitem{Aharonian:2004jr}
  F.~Aharonian and A.~Neronov,
 `High energy gamma--rays from the massive black hole in the galactic
center',
  {\it Astrophys. J.}  {\bf 619} (2005) 306
  [arXiv:astro-ph/0408303];
A.~Atoyan and C.~D.~Dermer,
`TeV emission from the galactic center black-hole plerion',
   {\it Astrophys. J.}  {\bf 617} (2004) L123
  [arXiv:astro-ph/0410243].






\end{thebibliography}
\bibliographystyle{unsrt}

\end{document}